\documentclass[twocolumn,showpacs,superscriptaddress,aps,prb,amsfonts,amsmath,amssymb,epsf,eqsecnum]{revtex4}
\usepackage[active]{srcltx}
\usepackage{epsfig}
\usepackage{amsfonts}
\usepackage{latexsym}
\usepackage{color}
\usepackage{epic}
\usepackage{amsfonts}
\usepackage[active]{srcltx}
\usepackage{epsfig}
\usepackage{latexsym}
\usepackage{epic}
\usepackage[colorlinks]{hyperref}

\newcommand{\ba}{\begin{array}}
\newcommand{\ea}{\end{array}}
\newcommand{\be}{\begin{equation}}
\newcommand{\ee}{\end{equation}}
\newcommand{\nn}{\nonumber}
\newcommand{\bea}{\begin{eqnarray}}
\newcommand{\ena}{\end{eqnarray}}
\newcommand{\beas}{\begin{eqnarray*}}
\newcommand{\enas}{\end{eqnarray*}}

\begin{document}

\title{Integrability of three dimensional models: cubic equations}

\author{Sh. Khachatryan}
\affiliation{Yerevan Physics Institute\\
Alikhanian Br. 2, 0036 Yerevan, Armenia}

\author{ A. Ferraz}
\affiliation{International Institute for Physics\\
Natal, Brazil}

\author{A. Kl\"umper}
\affiliation{Wuppertal University, Gau\ss{}stra\ss{}e 20, Germany}

\author{A. Sedrakyan}
\affiliation{Yerevan Physics Institute\\
Alikhanian Br. 2, 0036 Yerevan, Armenia}

\affiliation{International Institute for Physics\\
Natal, Brazil}


\date{\today}
\begin{abstract}
We extend basic properties of two dimensional integrable models within the
Algebraic Bethe Ansatz approach to 2+1 dimensions and formulate the sufficient
conditions for the commutativity of transfer matrices of different spectral
parameters, in analogy with Yang-Baxter or tetrahedron equations.  The basic
ingredient of our models is the R-matrix, which describes the scattering of a
pair of particles over another pair of particles, the quark-anti-quark (meson)
scattering on another quark-anti-quark state. We show that the Kitaev model
belongs to this class of models and its R-matrix fulfills well-defined equations
for integrability.


\end{abstract}

\maketitle

The importance of 2D integrable models \cite{Heisenberg, Bethe, Yang, Baxter,
  Faddeev} in modern physics is hard to overestimate.  Being initially an
attractive tool in mathematical physics they became an important technique in
low dimensional condensed matter physics, capable to reveal non-perturbative
aspects in many body systems with great potential of applications. The basic
constituent of 2D integrable systems is the commutativity of the evolution
operators, the transfer matrices of the models of different spectral
parameters. This property is equivalent to the existence of as many integrals
of motion as number of degrees of freedom of the model. It appears, that
commutativity of transfer matrices can be ensured by the Yang-Baxter (YB)
equations \cite{Yang,Baxter, Faddeev} for the R-matrix and the integrability of the
model is associated with the existence of the solution of YB-equations.

Since the 80s of last century there was a natural desire to extend the idea
of integrability to three dimensions \cite{Polyakov}, which resulted in a
formulation of the so-called tetrahedron equation by Zamolodchikov
\cite{Zamolodchikov}.  The tetrahedron equations (ZTE) were studied and several
solutions have been found until now \cite{Zamolodchikov, Bazhanov, Baxtertet,
  BaxtBazh, KashManStrog, KhS, Kuniba}.  However, earlier solutions either
contained negative Boltzmann weights or were slight deformations of models
describing free particles.  Only in a recent work \cite{ManBazhSer}
non-negative solutions of ZTE were obtained in a vertex formulation, and these
matrices can be served as Boltzmann weights for a 3D solvable model with
infinite number of discrete spins attached to the edges of the cubic
lattice. In this sense it is remarkable to note that among the general
solutions obtained in this paper it is also possible to detect R-matrices with
real and non-negative entries which can be considered as Boltzmann weights in
the context of the 3D statistical solvable models with 1/2-spins attached to
the vertexes of 3D cubic lattice.

Although initially the tetrahedron equations were formulated for the scattering
matrix S of three infinitely long straight strings in a context of 3D
integrability they can also be regarded as weight functions for statistical
models. In a Bethe Ansatz formulation of 3D models their 2D transfer matrices
of the quantum states on a plane \cite{Baxtertet,Bazhanov,KhS} can be
constructed via three particle R-matrix \cite{Hiet,Kor,Bazhanov}, which, as an
operator, acts on a tensorial cube of linear space V, i.e. $R : V \otimes V \otimes
V \rightarrow V \otimes V \otimes V$ \cite{footnote}.

Another approach to 3D integrability based on Frenkel-Moore simplex equations
\cite{IFGM} also uses three-state R-matrices. They are higher dimensional
extension of quantum Yang-Baxter equations without spectral parameters.
However these equations are less examined \cite{KGOSW}.

\begin{figure}[t]
\center
\centerline{\includegraphics[width=75mm,angle=0,clip]{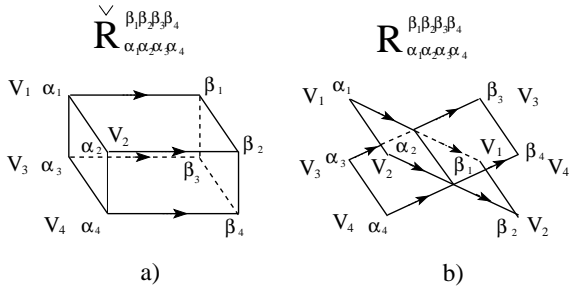}}
\caption{Four particle R-matrix}
\label{fig01}
\end{figure}

Motivated by the desire to extend the integrability conditions in
3D to other formulations we consider a new kind of equations with the
R-matrices acting on a quartic tensorial power of linear spaces V
\bea \label{R1} 
\check{R}_{1234} : V_1 \otimes V_2 \otimes V_3 \otimes V_4
\rightarrow V_1 \otimes V_2 \otimes V_3 \otimes V_4,
\ena 
which can be represented graphically as in Fig.\ref{fig01}a. The R-matrix can
be represented also in the form displayed in Fig.\ref{fig01}b, where the final
spaces are permuted ($V_1$ and $ V_2$ with $V_3$ and $V_4$, respectively):
$R_{1234}= \check{R}_{1234} P_{13}P_{24} $.  Explicitly it can be written as
follows
\bea
\label{R11}
R_{\alpha_1 \alpha_2 \alpha_3 \alpha_4}^{\beta_1 \beta_2 \beta_3 \beta_4}=
\check{R}_{\alpha_1 \alpha_2 \alpha_3 \alpha_4}^{\beta_3 \beta_4 \beta_1 \beta_2}.
\ena
Identifying
the space $V_1\otimes V_2$ and $V_3\otimes V_4$ with the quantum
spaces of quark-anti-quark pairs connected with a string one can
regard this R-matrix as a transfer matrix for a pair of
scattering mesons.
Within a terminology used in the algebraic Bethe Ansatz  for 1+1
integrable models this R-matrix  can be viewed also
as a matrix, which has two quantum states and two auxiliary states.

\begin{figure}[t]
\center
\centerline{\includegraphics[width=75mm,angle=0,clip]{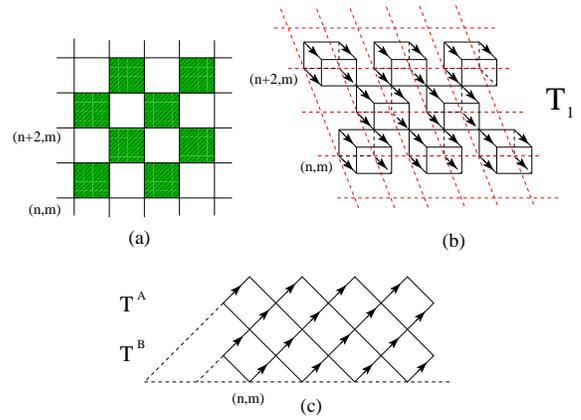}}
\caption{Transfer matrix T as a product of R-matrices on the plane: a)
  presents the lattice with chess like structure; b) presents the product of
  R-matrices arranged in chess like way; c) 2D cut of the product $T_B T_A$. }
\label{fig1}
\end{figure}

The space of quantum states $\Phi_t =\otimes_{(n,m)\in {\cal L}} V_{n,m}$ of
the system on a plane is defined by a direct product of linear spaces $ V_{n,m}$
of quantum states on each site $(n,m) $ of the lattice ${\cal L}$ (see
Fig.\ref{fig1}a). We fix periodic boundary conditions on both directions:
$V_{n,m+L}=V_{n,m}$ and $V_{n+L,m}=V_{n,m}$.  The time evolution of this
state is determined by the action of the operator/transfer matrix $T$:
$\Phi_{t+1}= \Phi_t T$, which is a product of local evolution operators,
R-matrices as follows. First we fix a chess like structure of squares on a
lattice ${\cal L}$ and associate to each of the black squares a R-matrix
$\check{R}_{(n+1,m)(n+1,m+1)(n,m)(n,m+1)}$, which acts on a product of four
spaces at the sites.  In this way the whole transfer matrix becomes
\bea
\label{R2}
T&=&Tr\Pi_{n=1}^{L/2} \Big[\Pi_{m=1}^{L/2}\check{R}_{(2n,2 m)(2n, 2m+1)(2n-1,2m)(2n-1, 2m+1)}\nn\\
 &\cdot& \Pi_{m=1}^{L/2} \check{R}_{(2n+1,2 m-1)(2n+1, 2m)(2n,2m-1)(2n, 2m)}\Big],
\ena
where the Trace is taken over states on boundaries. The indices of the R-matrices
in the first and second lines of this product just ensure chess like ordering
of their action.  In Fig.\ref{fig1}b we present this product
graphically. First we identify the second pair of states
$\langle(2n-1,2m),(2n-1,2m+1)\rangle$ (in first row) and
$\langle(2n+1,2m-1)(2n+1, 2m)\rangle$(in second row) of R-matrices with the
corresponding links on the lattice. Then we rotate the box of the R-matrix by
$\pi/4$ in order to ensure the correct order for their action in a product.
In the same way we define the second list of the transfer matrix, which will
act in the order $T_B T_A$. Fig.\ref{fig1}c presents a vertical 2D cut of two
lists of the product $T_B T_A$ drawn from the side. The $\pi/4$ rotated lines
mark the spaces $V_{n,m}$ attached to sites $(n,m)$ of the lattice.  Though
transfer matrix (\ref{R2}) is written in $\check{R}$ formalism, it can easily
be converted to the product of $R$-matrices.


The arrangement of R-matrices in the first row (first plane of the transfer
matrix $T_B$) acts on the sites of dark squares of the lattice while
R-matrices in the second row (second plane of the transfer matrix $T_A$) act
on the sites of the white squares.

Being an evolution operator the transfer matrix should be linked to
time. According to the general prescription \cite{Baxter, Faddeev} the transfer
matrix $T(u)$ is a function of the so-called spectral parameter $u$ and the linear
term $H_1$ in its expansion $T(u)=\sum_r u^r H_r $ defines the Hamiltonian of
the model, while the partition function is $Z= Tr T^N$. Integrable models
should have as many integrals of motion, as degrees of freedom. This property
may be reached by considering two planes of transfer matrices with different
spectral parameters, $T(u)$ and $T(v)$ and demanding their commutativity
$[T(u),T(v)]=0$, or equivalently demanding the commutativity of the
coefficients $[H_r, H_s]=0$ of the expansion.  This means, that all $H_r,\;
r>1$ are integrals of motion. In 2D integrable models the sufficient
conditions for commutativity of transfer matrices are determined by the
corresponding YB-equations \cite{Baxter,Yang, Faddeev}.

\begin{figure}[t]
\center
\centerline{\includegraphics[width=75mm,angle=0,clip]{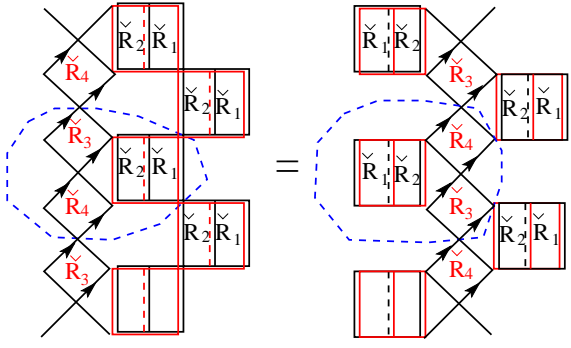}}
\caption{Reduced set of commutativity conditions for transfer matrices. The dotted
  subset represents the cubic equations (\ref{cub-2}). }
\label{fig11}
\end{figure}

In order to obtain the analog of the YB equations, which will ensure the
commutativity of transfer matrices (\ref{R2}) we use the so-called railway
construction. Let us cut horizontally two planes of the R-matrix product of two
transfer matrices (on Fig.\ref{fig1}b we present a product of R-matrices for
one transfer matrix plane) into two parts and substitute in between the
identity
\bea
\label{intertwiners}
&&\Pi_{m=1}^{L} id_{(2n+1, m)} id_{(2n, m)} \qquad \\
&=&\Big[Tr \Pi_{m=1}^{L}\check{\bar{R}}_{(2n+1, m)(2n+1, m+1)(2n, m)(2n, m+1)} \Big]^{-1}\nn\\
&\cdot&Tr\Big[\Pi_{m=1}^{L} \check{\bar{R}}_{(2n+1, m)(2n+1, m+1)(2n, m)(2n, m+1)} \Big]\nn
\ena
which maps two chains of sites, $(2n, m), m=1\cdots L $ and $ (2n+1, m),
m=1\cdots L+1$, into itself. The Trace have to be taken by identifying spaces
1 and L+1. In this expression we have introduced another set of
$\bar{R}$-matrices, called intertwiners, which will be specified below. For
further convenience we distinguish $\check{\bar{R}}_{(2n+1, m)(2n+1, m+1)(2n,
  m)(2n, m+1)}$ matrices for even and odd values of $m$ marking them as
$\check{R}_3$ and $\check{R}_4$ respectively. In the left side of
Fig.\ref{fig11} we present one half of the plane of R-matrices together with
an inserted chain of $\check{R}_3 \check{R}_4$ as intertwiners. The chain of
intertwiners can also be written by $R$-matrices.

Now let us suggest, that the product of these intertwiners with the first double
chain of $\check{R}$-matrices from the product of two planes of transfer
matrices is equal to the product of the same operators written in opposite
order. Namely we demand, that
\bea
\label{cub-1}
&&\Pi_{m=1}^{L}\check{\bar{R}}_{(2n+1, m)(2n+1, m+1)(2n, m)(2n, m+1)}\nn\\
&\cdot&\Pi_{m=1}^{L/2}\check{R}_{(2n,2 m)(2n, 2m+1)(2n-1,2m)(2n-1, 2m+1)}(u)\nn\\
 &\cdot& \Pi_{m=1}^{L/2} \check{R}_{(2n+1,2 m)(2n+1, 2m+1)(2n,2m)(2n, 2m+1)}(v)\nn\\
&=& \Pi_{m=1}^{L/2} \check{R}_{(2n,2 m)(2n, 2m+1)(2n-1,2m)(2n-1, 2m+1)}(v)\nn\\
 &\cdot& \Pi_{m=1}^{L/2} \check{R}_{(2n+1,2 m)(2n+1, 2m+1)(2n,2m)(2n, 2m+1)}(u)\nn\\
&\cdot&\Pi_{m=L}^{1}\check{\bar{R}}_{(2n+1, m)(2n+1, m+1)(2n, m)(2n, m+1)}. \qquad \qquad
\ena
Graphically this equation is depicted in Fig.\ref{fig11}. We move the column
of intertwiners from the left to the right hand side of the column of two
slices of the R-matrix product, simultaneously changing their order in a column,
changing the order of spectral parameters $u$ and $v$ of the slices and
demanding their equality. We can use the same type of equality and move the
chain of intertwiners further to the right hand side of the next column of the
two slices of the $\check{R}$-matrix product.  Then, repeating this operation
multiple times, one will approach the chain of inserted $\check{\bar{R}}^{-1}$
intertwiners inside the Trace from the other side and cancel it. As a result we
obtain the product of two transfer matrices in a reversed order of spectral
parameters $u$ and $v$.  Hence, the set of equations (\ref{cub-1}) ensures the
commutativity of transfer matrices.

\begin{widetext}

\begin{figure}[h]
\center
\centerline{\includegraphics[width=125mm,angle=0,clip]{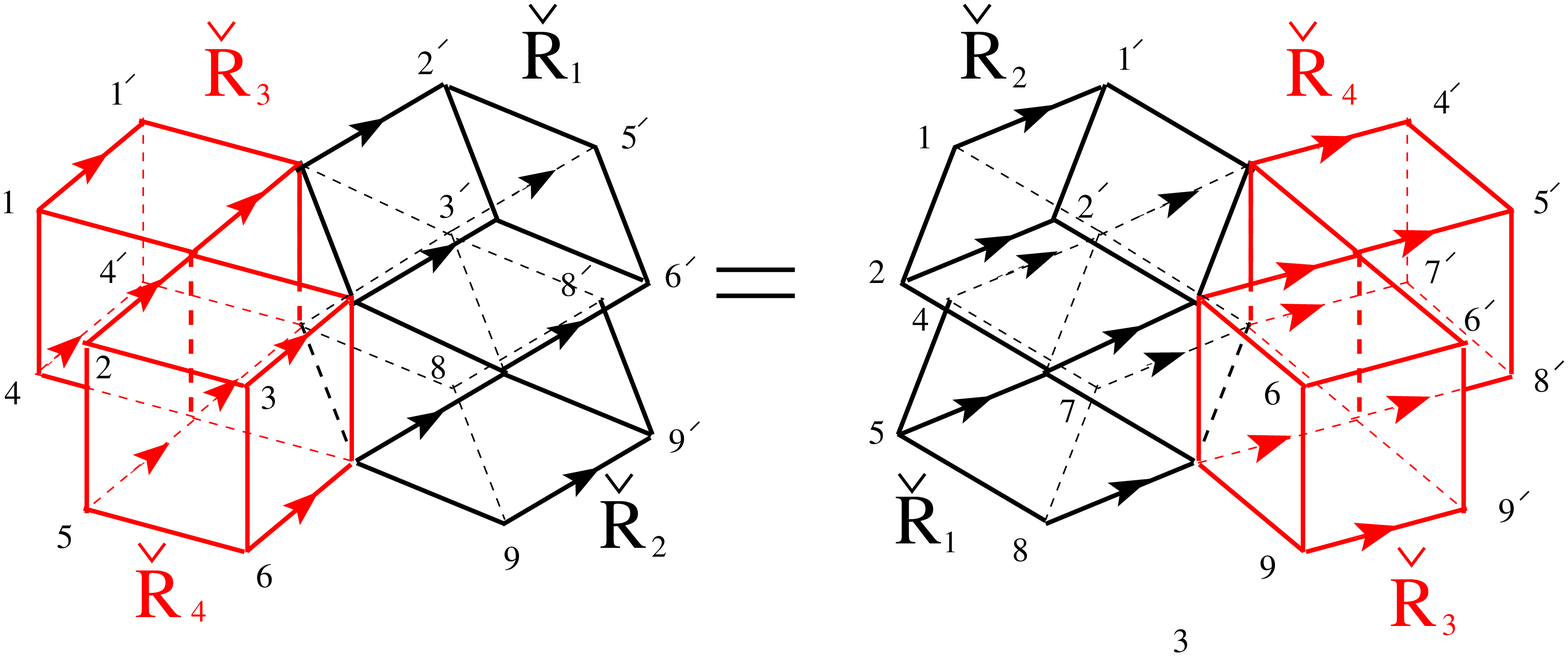}}
\caption{The set of equations ensuring commutativity of transfer matrices with
  different spectral parameters. We numerate linear spaces of states V, where
  R-matrices are acting, by $V_1 \cdots V_9$. R-matrices on the left hand side
  of equation are not acting on space $V_7$ while R-matrices on the right hand
  side are not acting on $V_3$. We put identity operators in the equation
  acting on this spaces for the consistency.  }
\label{fig2}
\end{figure}

\end{widetext}

The set of equations (\ref{cub-1}) can be simplified. Namely, it is easy to
see, that the equality can be reduced to the product of only 2
$\check{R}$-matrices, $\check{R}(u)$ and $\check{R}(v)$ and two intertwiners,
$\check{R}_3$ and $\check{R}_4$.  In other words, it is enough to write the
equality of the product of $\check{R}$-matrices from the inside of the dotted line
in Fig.\ref{fig11}.  Graphically this equation is depicted in
Fig.\ref{fig2}.

We see, that in this equation the product of $\check{R}$-matrices acting on a
space $\otimes_{i=1}^9 V_i$ (for simplicity we numerate the spaces from 1 to
9) can be written as
\bea
\label{cub-2}
&&\check{R}^4_{5263}(u,v)\check{R}^3_{4152}(u,v)\check{R}^2_{5689}(u)\check{R}^1_{2356}(v) id_7\nn \\
&&= \check{R}^1_{4578}(v)\check{R}^2_{1245}(u)\check{R}^3_{8596}(u,v)\check{R}^4_{7485}(u,v) id_3 \qquad
\ena
Here we have introduced a short-hand notation for $\check{R}$-matrices simply
by marking the numbers of linear spaces of states, in which they are acting;
$id_3$ and $id_7$ are identity operators acting on spaces 3 and 7
respectively.  Eq.~(\ref{cub-2}) can also easily be written by use of $R$.

This is the set of equations, sufficient for commutativity of transfer
matrices. The same set of equations are sufficient for commuting
$\check{R}$-matrices in the second column in Fig.\ref{fig01}.  Equations
(\ref{cub-2}) form an analog of YB equations ensuring the integrability of 3D
quantum models. Since they have a form of relations between the cubes of the
R-matrix picture(see Fig.\ref{fig01}) we call them cubic equations.

We will show now that the Kitaev model \cite{Kitaev} can be described as a
model of the prescribed type and its $R$-matrix fulfills the set of cubic
equations (\ref{cub-2}).  The full transfer matrix of Kitaev model is a
product $T_A T_B $ of two transfer matrices of type (\ref{R2}) defined by
$\check{R}$-matrices $\check{R}_A= 1 \otimes 1 \otimes 1 \otimes 1 + u\;
\sigma_x \otimes \sigma_x \otimes \sigma_x \otimes \sigma_x$ and $\check{R}_B=
1 \otimes 1 \otimes 1 \otimes 1 + u\; \sigma_z \otimes \sigma_z \otimes
\sigma_z \otimes \sigma_z$ respectively.  The linear term of the expansion of $T_A
T_B$ in the spectral parameter $u$ will produce the Kitaev model Hamiltonian
\bea
\label{Kit}
H_{Kitaev}&=&\sum_{white\ plaquettes} \sigma_x \otimes \sigma_x \otimes \sigma_x\otimes \sigma_x \nn\\
&+& \sum_{dark\ plaquettes} \sigma_z \otimes \sigma_z \otimes \sigma_z\otimes \sigma_z.
\ena

The integrability of the Kitaev model is trivially clear from the very
beginning since all terms in the Hamiltonian defined on white and dark
plaquettes commute with each other. The latter indicates, that the number of
integrals of motion of the model coincides with its degrees of freedom.
However, in this paper we are aiming to show, that one can develop 3D
Algebraic Bethe Ansatz approach in such a way, that the Kitaev model will
automatically be integrable. Namely, we will show now, that
$R_{A}$ and $R_{B}$-matrices of the Kitaev's model fulfill Eq.~\ref{cub-2}.
The explicit form of Eq.~\ref{cub-2} by use of indices according to the definition
in Fig.\ref{fig01}a reads
\bea 
&&\check{R^4}_{\alpha_5 \alpha_2 \alpha_6 \alpha_3}^{\beta_1 \beta_2 \beta_6 \beta_3}
\check{R^3}_{\alpha_4 \alpha_1 \beta_1 \beta_2}^{\gamma_4 \gamma_1 \beta_5 \beta_4}
\check{R^2}_{\beta_5 \beta_6 \alpha_8 \alpha_9}^{\beta_7 \beta_8 \gamma_8 \gamma_9}(v)
\check{R^1}_{\beta_4 \beta_3 \beta_7 \beta_8}^{\gamma_2 \gamma_3 \gamma_5 \gamma_6}(u)\delta_{\alpha_7}^{\gamma_7}=\nonumber\\
&&\check{R^1}_{\alpha_4 \alpha_5 \alpha_7 \alpha_8}^{\beta_7 \beta_8 \beta_5 \beta_6}(u)
\check{R^2}_{\alpha_1 \alpha_2 \beta_7 \beta_8}^{\gamma_1 \gamma_2 \beta_4 \beta_3}(v)
\check{R^3}_{\beta_6 \beta_3 \alpha_9 \alpha_6}^{\beta_1 \beta_2 \gamma_9 \gamma_6}
\check{R^4}_{\beta_5 \beta_4 \beta_1 \beta_2}^{\gamma_7 \gamma_4 \gamma_8 \gamma_5}\delta_{\alpha_3}^{\gamma_3}.\qquad \nonumber\\
 \label{check}
\ena
where $\check{R^1}(u)=\check{R_A}(u)$ and $\check{R^2}(v)=\check{R_B}(v)$.
It appears, that the intertwiners 
\bea
\label{sol}
\check{R^4}=\check{R_A}^{-1}(u),\;\;\check{R^3}=\check{R_B}(v) \qquad
\ena
where  $R_A^{-1}(u)=1 \otimes 1 \otimes 1 \otimes 1 - u \sigma_x \otimes \sigma_x \otimes \sigma_x \otimes \sigma_x$
fulfill the cubic equations (\ref{check}) for any parameters $u$ and $v$.
This can be directly checked both, by a computer algebra program and analytically. 
The commutativity of transfer matrices $T_A(u)$ with $T_A(v)$
and $T_B(u)$ with $T_B(v)$ is trivial in the Kitaev model. 
%

{\it Summary.} We have formulated a class of three dimensional models defined
by the $R$-matrix of the scattering of a two particle state on another two
particle state, i.e. a meson-meson type scattering. We derived a set of
equations for these $R$-matrices, which are a sufficient conditions for the
commutativity of the transfer matrices with different spectral
parameters. These equations differ from the tetrahedron equations, which also
ensure the integrability of 3D models, but are based on the R-matrix of 3 particle
scatterings.  Our set of equations will be reduced to tetrahedron type of
equations by considering the two auxiliary spaces in the R-matrix as one (fusion) and
replacing it by one thick line.  We showed that the Kitaev model \cite{Kitaev}
belongs to this class of integrable models.

{ \it Acknowledgment}.
A.S thanks IIP at Natal, where part of this work was done and Armenian Research Council (grant 13-1C132)
 for partial financial support.


\begin{thebibliography}{99}


\bibitem{Heisenberg}  W. Heisenberg, Z. Phys. 49, 619 (1928).

\bibitem{Bethe} H. Bethe, Z. Phys. 71, 205 (1931).

\bibitem{Yang}  C.N. Yang and C.P. Yang, Phys. Rev. 150, 321 (1966).

\bibitem{Baxter} R.J. Baxter, Ann. of Phys. 70, 193 (1972).

\bibitem{Faddeev} L.D. Faddeev, L.A. Takhtajan, Usp. Mat. Nauk, 34, 13 (1979) (in Russian).

\bibitem{Polyakov} A. Polyakov, 1979, unpublished.

\bibitem{Zamolodchikov}  A. Zamolodchikov, Zh. Eksp. Teor. Fiz. 79, 641 (1980). [English translation: Soviet Phys. JETP 52,
325 (1980)];  A. Zamolodchikov, Commun.Math.Physics 79, 489 (1981).

 \bibitem{Baxtertet} R.J.Baxter,
Commun. Math. Phys. 88, 185 (1983).

\bibitem{Hiet} J. Hietarinta, J. Phys. A: Math. Gen. 27, 5727, 5748 (1994).


\bibitem{BaxtBazh} V.V. Bazhanov, R.J.Baxter, J. Stat. Phys.
69 (1992) 453; V.V. Bazhanov, R.J.Baxter, Physica {\bf A 194}
(1993) 390-396.


\bibitem{footnote} Though in models with interaction round a cube
  \cite{Baxtertet,BaxtBazh} the basic R-matrix is defined on the product of four
  states: $V \otimes V \otimes V\otimes V$, which are situated on the vertices
  of a cube, however these models can be reformulated via the three channel
  R-matrix, where channels are associated with the faces of the cube.
  Therefore in this models also the equations of commutativity of transfer
  matrices are ZTE.

\bibitem{Kitaev} A.Yu. Kitaev, Annals of Physics 303, 2 (2003).




\bibitem{KashManStrog} R.M. Kashaev, V.V. Mangazeev, Yu.G. Stroganov, Int. J. Mod. Phys.
A8 (1993) 587-601;\;
 R.M. Kashaev, V.V. Mangazeev, Yu.G. Stroganov, Int. J. Mod. Phys.
{\bf A8} (1993) 1399-1409.






\bibitem{Bazhanov} V.V. Bazhanov, V. V. Mangazeev, S.M. Sergeev,
 J. Stat. Mech. P07004 (2008);\;
V.V. Bazhanov, S.M. Sergeev,  J. Phys. A: Math. Theor. 39
3295-3310 (2006).

\bibitem{ManBazhSer}  V. V. Mangazeev, V.V. Bazhanov, S.M. Sergeev, J.
Phys. A: Math. Theor. {\bf 46} 465206 (2013).


\bibitem{GePaSer}  G. von Gehlen, S. Pakuliak and S. Sergeev, 
 J. Phys. A: Math. Gen. {\bf 36},975 (2003);\; G. von Gehlen, S. Pakuliak and S. Sergeev,
  Int. J. Mod. Phys. {\bf A 19} Suppl. 179-204 (2004) ; \;
G. von Gehlen, S. Pakuliak, S. Sergeev, J.Phys. {\bf A 38},
7269  (2005).

\bibitem{KhS} J. Ambjorn, Sh. Khachatryan, A. Sedrakyan, Nucl. Phys. B 734, [FS] , 287 (2006).

\bibitem{IsKul} A. P. Isaev and P. P. Kulish, Mod. Phys. Lett. {\bf A 12} 427
(1997).

\bibitem{Kuniba} A. Kuniba, M. Okado, J. Phys. A: Math. Theor. {\bf 45} 465206,(2012). 

\bibitem{ZamAlg} I.G. Korepanov, 
Modern Phys. Lett. {\bf B 3}, No 3 (1989) 201-206.

\bibitem{Kor} I.G. Korepanov, Comm. Math. Phys. 154 (1993)
85.


\bibitem{KhSZ} Sh. Khachatryan, A. Sedrakyan, J. Stat. Phys. {\bf 150} 130
(2013).

\bibitem{IFGM}  I. Frenkel and G. Moore, Commun. Math. Phys. {\bf 138} (1991) 259.

\bibitem{KGOSW}
 M.~L.~Ge, C.~H.~Oh and K.~Singh, Phys. Lett. A 185, 177  (1994);\;
  L.~C.~Kwek, C.~H.~Oh, K.~Singh and K.~Y.~Wee,
J. Phys. A: Math. Gen. {\bf 28}, 6877 (1995).

\end{thebibliography}
\end{document}